\documentclass{article}
\usepackage{makeidx}
\usepackage{graphicx}
\usepackage{amsmath}

\input{tcilatex}

\begin{document}

\title{Quantum games with a multi-slit electron diffraction setup }
\author{A. Iqbal \\
Department of Electronics, Quaid-i-Azam University, \\
Islamabad 45320, Pakistan\\
email: qubit@isb.paknet.com.pk\\
}
\maketitle

\begin{abstract}
A setup is proposed to play a quantum version of the famous bimatrix game of
Prisoners' Dilemma. Multi-slit electron diffraction with each player's pure
strategy consisting of opening one of the two slits at his/her disposal are
essential features of the setup. Instead of entanglement the association of
waves with travelling material objects is suggested as another resource to
play quantum games.
\end{abstract}

\section{Introduction}

Quantum games has become the subject of many recent interesting papers \cite
{meyer,eisert,benjamin,eisert1,marinatto,du,iqbal}. Meyer, in a seminal work 
\cite{meyer}, presented the idea of playing a sequential quantum game with
unitary manipulation of a qubit. Shortly afterwards Eisert, Wilkens, and
Lewenstein \cite{eisert,eisert1} aroused a lot of interest in quantum games
by showing that the dilemma in the famous classical game of Prisoners'
Dilemma (PD) can disappear when the game is played with quantum mechanical
maneuvers consisting of local and unitary manipulation of a pair of
maximally entangled qubits. Much of the later analysis of quantum games
consists of exploiting the popular resource of quantum information, i.e. the
entanglement, in many different proposed setups. Many authors have made
their valuable contributions by providing many interesting scenarios for
quantum games \cite{Ariano,piotrowski,lung,grib,doescher,lee}. Unitary
manipulation of entangled systems of qubits is one of the favorite general
concept to play a quantum game \cite
{meyer,eisert,benjamin,marinatto,du,iqbal}. Though it is shown that
entanglement may not be essential for a quantum game \cite{du1} but it
continues to play its role as a resource. A noticeably greater attention
paid to many suggestions exploiting entanglement for quantum games can be
traced back to the exciting and counter-intuitive properties of this
phenomenon, as well as to its recent active and detailed investigation in
quantum information theory \cite{nielsen}.

Unitary manipulation of entangled qubits to play matrix games is indeed an
interesting concept giving new dimensions to the classical game theory. But
it doesn't forbid the use of other quantum mechanical effects to play other
`quantum forms' of the matrix games---games for which extensive classical
analysis is already found in literature \cite{burger,gibbons}. A look at the
Eisert et al.'s idea \cite{eisert,eisert1} makes apparent some of its
similarities to the well known Young's double-slit apparatus \cite{hecht}.
Simultaneous and local unitary manipulation of a maximally entangled
two-qubit quantum state and its later measurement is the essential feature
of Eisert et al.'s idea. In Young's double-slit case, however, a coherent
light passes through two slits to form a diffraction pattern on a screen
facing the slits. Similarity between these setups becomes noticeable if one
draws a kind of comparison between the properties of entanglement and
coherence, players' moves and the process of opening or closing the slits,
wavefunction-collapsing measurement and the appearance of the diffraction
pattern, in the setups of Eisert et al. and Young respectively. This
comparison in its turn asks for some quantum feature that can be exploited
to give some new dimension to a matrix game when played using a Young's
double-slit like apparatus. In Eisert et al.'s setup this quantum feature is
obviously the entanglement. In Young's apparatus this feature can be, for
example, the association of well known wave-like properties to material
objects like electrons, producing a diffraction pattern on a screen.

Historically speaking the De Broglie's original idea \cite{hecht,debroglie},
that travelling material particles have waves associated with them, was
undoubtedly the key concept behind the development of quantum physics in
early part of the twentieth century. Soon afterwards Davisson and Germer 
\cite{hecht} verified the idea in their experimental demonstration of the
diffraction of electrons by crystals. In De Broglie's argument a travelling
electron with momentum $p$ has an associated wave of wavelength $\lambda
=h/p $, where $h$ is the Plank's constant. To make $\lambda $ a measurable
quantity, under normal laboratory conditions, the momentum $p$ should have
similar order of magnitude as $h$. It shows why it is very hard to detect
waves associated with macroscopic objects. Our motivation is to take the
quantum feature, that associates wave-like properties with micro objects, as
a resource that can be used to play a quantum game. Such a quantum game can
be realized using an apparatus consisting of travelling electrons, multiple
slits intercepting them, and a resulting diffraction pattern. In this setup
a player's choice of a `pure strategy' will consist of opening or closing
slits at his/her disposal. Suppose the apparatus is adjusted such that when $%
\lambda $ approaches zero the classical game is reproduced. It can then be
argued that because an observation of a value of $\lambda $ quite away from
zero is entirely a quantum feature, therefore, the resulting different
payoffs for the players correspond to a quantum form of the classical game.
In this setup the players' payoffs are to be found from the diffraction
pattern formed on the screen. We show the possibility of finding a value for 
$\lambda $ that makes appear a non-classical equilibrium in the PD game when
the players play only the pure strategies. The classical game remains a
subset of its quantum version because with $\lambda $ approaching zero the
classical game is reproduced.

\section{Playing quantum games with a diffraction apparatus}

Eisert et al.'s first investigation of quantum Prisoners' Dilemma (PD) \cite
{eisert,eisert1} has provided much of the later motivation for a systematic
study of quantum games. The classical game of PD, in its general form, can
be represented by the matrix in the fig.(\ref{fig.1}).

\FRAME{ftbpFU}{2.8885in}{1.177in}{0pt}{\Qcb{Payoff matrix for general
Prisoners' Dilemma. The first and the second entries in a parenthesis are
Alice's and Bob's payoffs, respectively. For Prisoners' Dilemma the
condition $t>r>p>s$ should hold (See in Ref. \protect\cite{du2, straffin}).}%
}{\Qlb{fig.1}}{pdilemma.eps}{\special{language "Scientific Word";type
"GRAPHIC";maintain-aspect-ratio TRUE;display "USEDEF";valid_file "F";width
2.8885in;height 1.177in;depth 0pt;original-width 3.8199in;original-height
1.6994in;cropleft "0.0362";croptop "0.9174";cropright "0.9611";cropbottom
"0.0822";filename '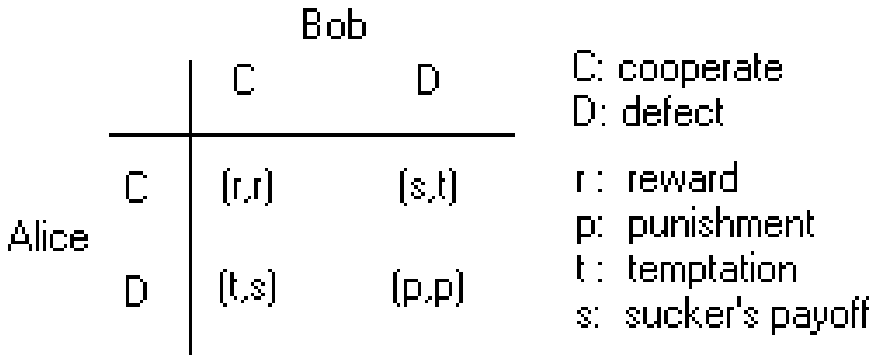';file-properties "XNPEU";}}Our motivation to
play a quantum form of PD game, without using the phenomenon of
entanglement, derives from Feynman's excellent exposition \cite{feynman} of
quantum behavior of atomic objects. He describes and compares the
diffraction patterns---in two similar imaginary, but experimentally
realizable, setups---consisting of bullets and electrons passing through a
wall with two slits. Feynman describes a well known quantum
property---associating waves to all material particles---to distinguish the
diffraction patterns of bullets and electrons. The disappearance of a
pattern for the bullets is then explained as due to tiny wavelengths of the
associated waves, such that the pattern becomes very fine, and with a
detector of finite size one can not distinguish the separate maxima and
minima. We asked why not to play a game, in the Feynman's imaginary
experimental setup, such that one gets the classical game when, in Feynman's
words, bullets are fired and a quantum game correspond when electrons
replace the bullets. The experimental setup shown in the fig.(\ref{fig.2})
illustrates the point.

\FRAME{ftbpFU}{3.1324in}{2.0954in}{0pt}{\Qcb{A multi-slit diffraction setup
to play a quantum form of Prisoners' Dilemma. A window with four slits faces
an electron source. Each player has access to two slits. A player plays a
pure strategy by opening a slit and closing the other. Arbiter finds the
players' payoff by measuring the peak-to-peak distance on the diffraction
pattern formed on the screen.}}{\Qlb{fig.2}}{diffrgame.eps}{\special%
{language "Scientific Word";type "GRAPHIC";maintain-aspect-ratio
TRUE;display "USEDEF";valid_file "F";width 3.1324in;height 2.0954in;depth
0pt;original-width 4.7798in;original-height 3.4541in;cropleft
"0.0352";croptop "0.9044";cropright "0.9633";cropbottom "0.0498";filename
'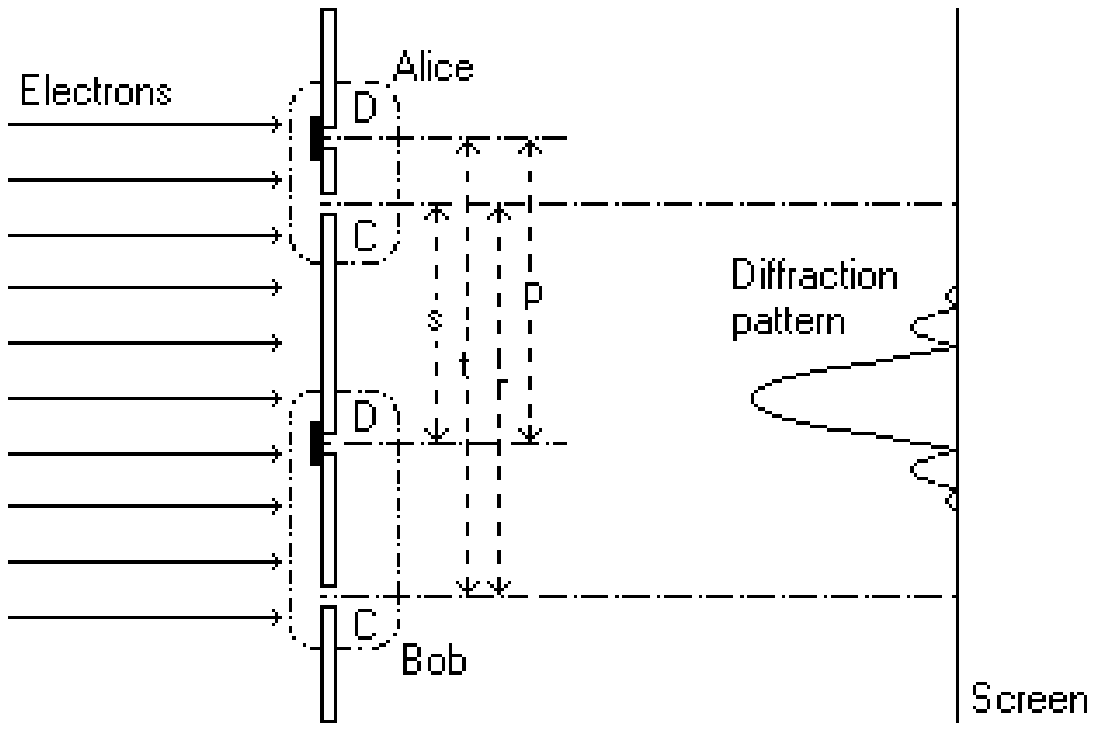';file-properties "XNPEU";}}

We now select the famous classical PD game to be played in this setup. To
make the classical game imbedded in its quantum version the positive
coefficients $p,r,s,$ and $t$, appearing in the matrix representation of the
game, are translated into the distances between the slits. Each player is in
control of two slits such that his/her strategy consists of opening one of
the slits and closing the other. For example, if Alice decides to cooperate
then she would open the slit $C$ and close $D$. Because Bob has a similar
choice, therefore, every possible move by the players leads to opening of
two slits and closure of the other two, with the separation between the two
open slits depending on the moves of the players. It will happen when only
the so-called `pure-strategies' can be played by the players, which in the
present setup means to open a slit and close the other. Now, at the final
stage of the game, the action of the arbiter---who is responsible for
computing the payoffs when the players have made their moves---consists of
measuring the distance between two peaks of the diffraction pattern. This
peak-to-peak distance is known to be $\lambda /d$ \cite{hecht}, where $d$ is
the separation between the open sits and $\lambda $ is the wavelength
associated with the bombarded material objects, like electrons. The payoffs
for the players are the functions of $\lambda /d$ and it, then, explains the
utility of translating the coefficients of the matrix of the classical game
into separations $d$ between the slits. When bullets are fired, which means
here the particles become heavier and corresponding $\lambda $ is very
nearly zero, the payoffs become classical and depend only on $d$ i.e. the
separation between the slits. A payoff representation in terms of $\lambda
/d $ contains both the classical and quantum aspects of the matrix game
played in this setup.

For PD the payoffs are symmetric for the players and a single equation can
describe the payoffs to both the players when their strategies are known. A
usual way to express it is to write $P(s_{1},s_{2})$ for the payoff to the $%
s_{1}$-player against the $s_{2}$-player. Such a single equation
representation is usually used in evolutionary games \cite{weibull}
consisting of symmetric bimatrix conflicts. The $s_{1}$-player is referred
to as the `focal' player and the $s_{2}$-player as just the `other' player.
The PD is one such example for which a single payoff equation can capture
the essence of the idea of a symmetric Nash equilibrium (NE). The strategy $%
s^{\star }$ is symmetric NE if

\begin{equation}
P(s^{\star },s^{\star })-P(s,s^{\star })\geq 0,\text{ \ \ \ for all }s\neq
s^{\star }  \label{NE}
\end{equation}
saying that the focal player can not be better off by diverging away from $%
s^{\star }$. Because the setup of the fig.(\ref{fig.2}) involves only the
coefficients in the classical payoff matrix corresponding to the first
player, therefore, finding a symmetric NE with eq.(\ref{NE}) becomes
possible immediately when the first player is taken as focal. It also shows
why writing payoff as $P(s_{1},s_{2})$ is relevant to setup of the fig.(\ref
{fig.2}). For example, classically the strategy of defection $D$ comes out
as a symmetric NE because $P(D,D)-P(C,D)=(p-s)>0$, when the players' moves
consist of the pure strategies only.

In the setup of the fig.(\ref{fig.2}), for every pure strategy move the
players have option to make, a unique separation $d$ between the slits is
obtained that can have four possible values i.e. $p,r,s$ or $t$. Classically 
$P(C,C)=r$, $P(C,D)=s$, $P(D,C)=t$, and $P(D,D)=p$. It can be noticed in the
fig.(\ref{fig.2}) that the classical payoff to the focal player against the
other can be equated to the separation between the two open slits $d$.

Now assume that the arbiter uses the following payoff equation, instead of
simply $P(s_{1},s_{2})=d$.

\begin{equation}
P(s_{1},s_{2})=d+k(\lambda /d)
\end{equation}
where $k$ is a positive constant that can be called a scaling factor. $%
P(s_{1},s_{2})$, obviously, reduces to its classical counterpart when $%
\lambda $ is very nearly zero. Suppose the strategy of cooperation $C$ is a
symmetric NE then

\begin{equation}
P(C,C)-P(D,C)=\left\{ k\lambda (1/r-1/t)-(t-r)\right\} \geq 0
\end{equation}
It requires $\lambda \geq rt/k$. For electrons of mass $m$ travelling with
velocity $v$ it gives $v\leq (kh/mrt).$ Supposing $r$ and $t$ are both non
zero, the arbiter's problem consists of finding an appropriate value for the
scaling factor $k$ for which $v$ comes in a reasonable range from
experimental point of view. When the electrons have a $\lambda \geq rt/k$
the strategy of cooperation becomes a symmetric NE and each player gets a
payoff $r+k\lambda /r$. Similarly when the pure strategy of defection is a
symmetric NE in the quantum game

\begin{equation}
P(D,D)-P(C,D)=\left\{ -k\lambda (1/s-1/p)+(p-s)\right\} \geq 0
\end{equation}
It requires $\lambda \leq sp/k$. After the scaling factor $k$ is determined
the wavelength $\lambda $ decides which pure strategy should be a symmetric
NE. Two ranges for $\lambda $ are, therefore, indicated i.e. $\lambda \leq
sp/k$ and $\lambda \geq rt/k$. Defection and cooperation come up as
symmetric NE for these ranges, respectively. Because the constants $t,r,p,s$
and $k$ are all positive the classical game is in the earlier range of $%
\lambda $. Non-classical equilibrium of cooperation shows itself in the
later range of $\lambda $.

Du et al.'s recent analysis \cite{du} of the quantum PD, with players'
access to Eisert's set of two-parameter set of unitary operators, has shown
an intriguing structure in the game as a function of the amount of
entanglement. The game turns into classical with the amount of entanglement
becoming zero. In the setup of fig.(\ref{fig.2}) the quantity $\lambda $
behaves in similar way like the amount of entanglement in Du et al.'s
analysis \cite{du,du2}. But the present setup, to play a quantum game, is
devoid of the notion of entanglement and relies instead on a very different
quantum aspect, which is as much quantum in nature as the phenomenon of
entanglement for qubit systems. There is however a difference, to be
noticed, between the setups of Eisert et al. and of the fig.(\ref{fig.2}).
Players' actions in Eisert et al.'s setup are very much quantum mechanical
in nature in that they make moves with operators from the quantum world. In
the present setup, on the contrary, the players' actions are entirely
classical consisting of opening or closing slits. It is similar to the
players' actions in Marinatto and Weber's idea \cite{marinatto} of playing a
quantum form of the matrix game of the battle of sexes. In this scheme
players possess quantum operators but they apply those on an initial quantum
state with classical probabilities, so that the players' moves can be
considered classical as well. A transition to the classical game is obtained
by unentangling the initial state. It can be observed that, apart from the
pioneering work of Eisert et al., the setup of fig.(\ref{fig.2}) is also
motivated, to an almost equal extent, by the Marinatto and Weber's idea \cite
{marinatto} of playing a quantum version of a matrix game.

\section{Concluding remarks}

In many of the earlier suggested setups some appropriate measure of
entanglement for a qubit system is introduced. The quantum version of the
game reduces to classical when the measure becomes zero. Instead of the
entanglement we introduce another resource from quantum physics, i.e. the
association of waves with travelling material objects like electrons, to
show how it can lead to a non-classical equilibrium in the PD game. With
associating wavelength approaching zero the quantum aspect disappears and
the classical game is reproduced. Like entanglement the association of waves
with particles is a purely quantum phenomenon rightly considered a
cornerstone of the quantum theory, though less esoteric because of its
earlier discovery. We suggest its exploitation as another resource to play
quantum games.

\end{document}